\documentclass{emulateapj}

\newcommand{\etal}{et al.}
\newcommand{\ezages}{EZ\_Ages }

\newbox\grsign \setbox\grsign=\hbox{$>$} \newdimen\grdimen \grdimen=\ht\grsign
\newbox\simlessbox \newbox\simgreatbox
\setbox\simgreatbox=\hbox{\raise.5ex\hbox{$>$}\llap
     {\lower.5ex\hbox{$\sim$}}}\ht1=\grdimen\dp1=0pt
\setbox\simlessbox=\hbox{\raise.5ex\hbox{$<$}\llap
     {\lower.5ex\hbox{$\sim$}}}\ht2=\grdimen\dp2=0pt
\def\simgreater{\mathrel{\copy\simgreatbox}}
\def\simless{\mathrel{\copy\simlessbox}}
\newbox\simppropto
\setbox\simppropto=\hbox{\raise.5ex\hbox{$\sim$}\llap
     {\lower.5ex\hbox{$\propto$}}}\ht2=\grdimen\dp2=0pt

\slugcomment{To appear in The Astronomical Journal}

\shorttitle{M~31 Globular Clusters}
\shortauthors{Schiavon \etal}

\begin{document}



\title{Star Clusters in M~31. V. Evidence for Self-Enrichment in
Old M31 Clusters from Integrated Spectroscopy}


\author{Ricardo P. Schiavon}
\affil{Astrophysics Research Institute, 
Liverpool John Moores University, 146 Brownlow Hill, Liverpool, L3
5RF, United Kingdom.}
\email{R.P.Schiavon@ljmu.ac.uk}
\author{Nelson Caldwell}
\affil{Harvard-Smithsonian Center for Astrophysics, 60 Garden Street,
Cambridge, MA 02138, USA}
%\email{caldwell@cfa.harvard.edu}
\author{Charlie Conroy}
\affil{Department of Astronomy \& Astrophysics, University of California,
Santa Cruz, CA, USA}
%\email{cconroy@ucolick.org}
\author{Genevieve J. Graves}
\affil{Department of Astrophysical Sciences, Princeton University,
Princeton, NJ, USA}
%\email{graves@astro.princeton.edu}
\author{Jay Strader}
\affil{Department of Physics and Astronomy, Michigan State University,
East Lansing, MI 48824, USA}
%\email{strader@pa.msu.edu}
\author{Lauren A. MacArthur}
\affil{National Research Council of Canada, Victoria, BC V9E 2E7, Canada}
%\email{Lauren.MacArthur@nrc-cnrc.gc.ca}
\author{St\'ephane Courteau}
\affil{Department of Physics, Engineering Physics \& Astronomy, Queen’s University,
Kingston, Ontario, Canada}
%\email{courteau@astro.queensu.ca}
\and
\author{Paul Harding}
\affil{Department of Astronomy, Case Western Reserve University, Cleveland, OH 44106-7215, USA}
%\email{paul.harding@case.edu}



\begin{abstract}

In the past decade, the notion that globular clusters (GCs) are
composed of coeval stars with homogeneous initial chemical compositions
has been challenged by growing evidence that they host an intricate
stellar population mix, likely indicative of a complex history of
star formation and chemical enrichment.  Several models have been
proposed to explain the existence of multiple stellar populations
in GCs, but no single model provides a fully satisfactory match to
existing data.  Correlations between chemistry and global parameters
such as cluster mass or luminosity are fundamental clues to the
physics of GC formation.  In this {\it Letter}, we present an
analysis of the mean abundances of Fe, Mg, C, N, and Ca for 72 old
GCs from the Andromeda galaxy.  We show for the first time that
there is a correlation between the masses of GCs and the mean stellar
abundances of nitrogen, spanning almost two decades in mass.  This
result sheds new light on the formation of GCs, providing important
constraints on their internal chemical evolution and mass loss
history.

\end{abstract}



\section{Introduction}

Globular clusters (GCs) are the oldest known stellar systems and
their study has contributed fundamental knowledge to our understanding
of stellar evolution and galaxy formation \citep[e.g.,][]{sz78,az92,bs06}.
Yet their origin remains poorly understood.  The discovery of
multiple stellar populations with a range of chemical compositions
in Milky Way (MW) GCs revolutionized our understanding of GC
formation, as these systems were previously believed to form in a
single rapid star formation event.  Several scenarios were proposed
to explain the observed complexities of the stellar populations in
GCs \citep[e.g.,][]{re08,de10,de07,dm09,ma09,vc11,cs11}, but all
fail to match all the observations of Galactic GCs.  In common to
most models is the following triad of features: 1) GCs underwent
internal chemical evolution;  2) more massive GCs self-enriched
more efficiently due to an increased ability to hold on to enriched
gas for subsequent star formation;  and 3) GCs have lost {\it most}
of their primordial mass.

It is difficult to use resolved observations of Galactic GCs to
submit the above predictions to quantitative scrutiny, as this
requires the determination of detailed abundance patterns for
statistically significant samples of individual GC stars spanning
a wide range of GC masses.  Such determinations are very costly,
as they rely on high-resolution spectroscopy based on 8-10~m
telescopes.  Integrated spectroscopy can provide important supplementary
information to constrain models, in the form of mean abundances of
elements such as nitrogen, which indicate the presence of self-enrichment
in GCs.  Based on a sample of 72 old GCs belonging to the Andromeda
(M31) galaxy we study the correlations between GC mass and the mean
abundances of Fe, Mg, C, N, and Ca.  For the first time, we
find evidence for the presence of a correlation between GC mass and
the mean abundance of nitrogen.  This result challenges the notion
that GCs have lost most of their masses since formation, and may
help constrain the source of chemical enrichment in these systems.

This {\it Letter} is organized as follows: the data are described
in Section~\ref{data}, the results are presented in Section~\ref{results},
and our conclusions are summarized in Section~\ref{summary}.

\section{Data} \label{data}

The M31 GC spectra employed in this analysis \citep{ca09} were obtained
with the MMT/Hectospec spectrograph, at Mount Hopkins, Arizona,
whereas the spectra of MW GCs were collected with the Boller \& Chivens
spectrograph attached to the Blanco 4~m telescope at Cerro Tololo
Interamerican Observatory \citep{s05}.  Absorption line indices,
measured in the Lick system, as well as internal and systematic
uncertainties, are discussed in detail elsewhere \citep{s12}.

\subsection{Elemental Abundances} \label{abunds}

Lick index measurements are converted into abundance and age
estimates, along with associated uncertainties, using the \ezages
code \citep{gs08}, which implements a method developed by \cite{s07}.
For more details on abundance error estimates, including extensive
Monte Carlo simulations, see \cite{gs08}.  Nitrogen abundances are
based on the CN$_1$ and/or CN$_2$ indices, and because the CN band
measured by these indices is too weak to be detected in our medium
resolution integrated spectra of GCs with [Fe/H] $\simless$ --1.0
\citep{s12}, only GCs more metal-rich than that limit are considered.
Thus, our initial sample includes 175 GCs with iron abundance within
the range --1.0 $\le$ [Fe/H] $\le$ +0.2.  The sample is further
restricted to GCs for which our spectra have S/N$\geq$100/${\rm\AA}$,
for which our abundances, particularly for nitrogen, are most
reliable.  The final sub-sample of 72 GCs spans masses between
10$^{4.8}$ and 10$^{6.5}$ $M_\odot$.  \cite{ca09} and \cite{s12}
showed that M31 GCs are about as old as their MW couterparts,
thus the two GC systems can be compared on the same footing.

%The procedure followed to derive abundance errors was described in
%detail by \cite{gs08}, so we only summarize it briefly here.
%Abundance errors are based on a propagation of the errors in line
%indices and stellar population parameters.  For instance, the error
%in the abundance of nitrogen is determined by the error in the index
%used in its determination (e.g., CN$_1$) and the errors in parameters
%upon which that index is also dependent, such as age, [Fe/H], and
%[C/Fe].  Extensive Monte Carlo simulations showed that the \ezages
%errors are reliable in most cases, except perhaps at low S/N, where
%\ezages slightly overestimates the errors, but given the S/N of the
%sample under analysis, this is irrelevant.

\subsection{Cluster Masses}

Mass is a key parameter in our analysis, and we note that elemental
abundances and masses are based on independent data and methods.
Within the sample of 72 GCs studied in this {\it Letter}, dynamical
masses ($M_{dyn}$) from \cite{st11} are available for 53 objects
and, for the remaining GCs, photometric masses ($M_{phot}$) are
adopted, estimated assuming $m-M = 24.43$ and M/L$_V$=1.5 or
M/L$_K$=0.8, depending on available photometry.  Masses for 
MW GCs were taken from \cite{mm05} and \cite{go97}.

\begin{figure}
\includegraphics[scale=0.45]{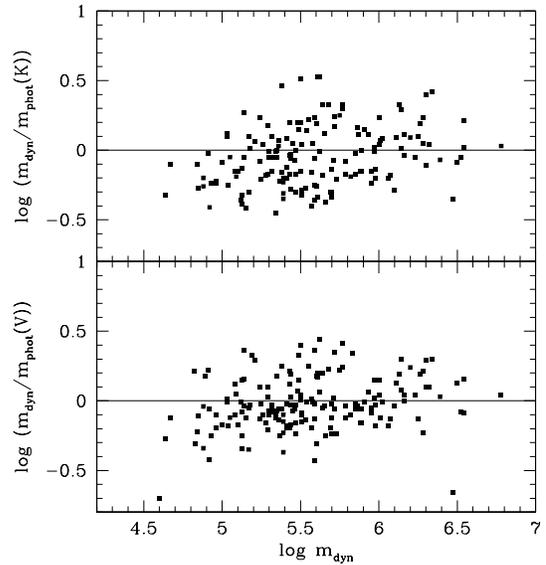}
\caption{Comparison between dynamical and photometric masses, based
on K- (top panel) or V-band (bottom panel) magnitudes.  The three
sets of GC masses are in very good agreement, without significant
zero point differences.
} 
\label{fig1}
\end{figure}

\begin{figure} 
\includegraphics[scale=0.45]{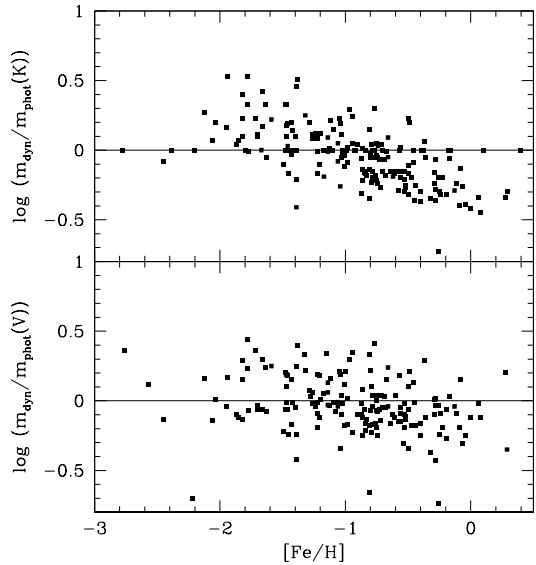}
\caption{Comparison between dynamical and photometric masses, as a
function of [Fe/H].  Photometric masses are slightly smaller at
higher [Fe/H], but the effect is negligible.
}
\label{fig2}
\end{figure}

We next check for the presence of systematic effects on $M_{phot}$
due to the impact of age and/or chemical composition on mass-to-light
ratios.  In Figures~\ref{fig1} and \ref{fig2}, V band- and K
band-based $M_{phot}$s are compared with $M_{dyn}$ as a function
of $M_{dyn}$ and [Fe/H], for a larger GC sample from \cite{ca09}.
Both sets of photometric masses are in good agreement with $M_{dyn}$
and in Figure~\ref{fig1} the residuals show at most a very slight
correlation with GC mass, which is negligible for the purposes of
this {\it Letter}.  In Figure~\ref{fig2} a trend can be seen with
[Fe/H], particularly in the case of K band-based photometric masses.
The effect is very small, however, especially considering the [Fe/H]
range of the sample ($-1.0 \leq {\rm [Fe/H]} \leq +0.2$), and the
fact that photometric masses were adopted for only $\sim$1/4 of
those GCs.  Therefore, systematics on the K band-based photometric
masses is at most of the order of 0.2 dex, and half of that in the
case of V band-based masses.  Considering the dynamic range (in
mass) of our correlations, this effect is negligible.  In fact, the
statistics on the correlations studied in this {\it Letter} is
unaffected regardless of whether we employ only V-band or K-band
photometric masses.  We conclude that our analysis incurs no important
systematic effects due to the adoption of photometric masses for
part of the sample.  For an in depth discussion of Figures~\ref{fig1}
and \ref{fig2}, see \cite{st11}.

\section{Results}  \label{results}

\begin{figure} 
\includegraphics[scale=0.45]{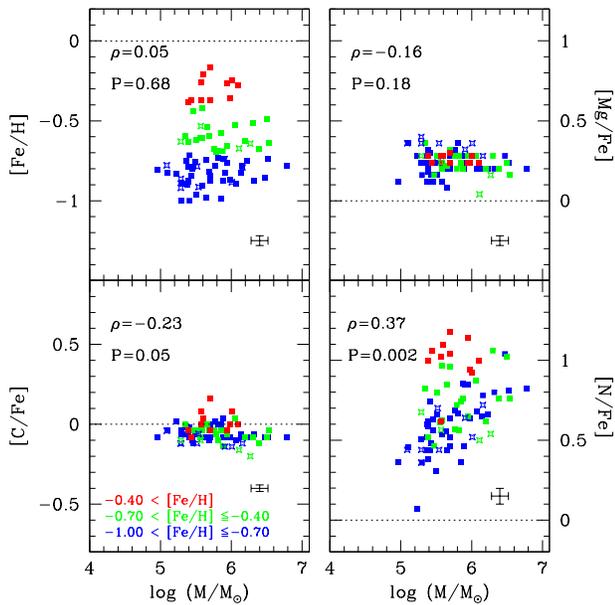}
\caption{
Globular clusters are displayed in several mass-chemistry planes.
Mean iron abundance and abundance ratios are plotted for M31 (gray
squares) and MW GCs (stars).  Vertical scales span the same dynamic
range in all panels, for an immediate visual assessment of the
relative strength of correlations and amount of scatter in the data.
Average error bars are displayed on the lower right corner of each panel.
Dotted lines indicate the solar value.  The Spearman rank correlation
coefficient for each pair, and the probability associated with the
null hypothesis are displayed on the top left corner of each panel.
Our main result is the correlation between [N/Fe] and GC mass (bottom
right panel), which is consistent with the presence of self-enrichment
in M31 GCs.  The correlation is more obvious at fixed [Fe/H], due
to a concurrent correlation between [N/Fe] and [Fe/H] (see
Figure~\ref{fig4}).  The absence of a correlation in heavy elements
such as Fe, Mg, and Ca indicates that N enrichment cannot
be caused by explosive nucleosynthesis.  The numbers suggest the presence
of a weak anti-correlation between [C/Fe] and mass, which is not confirmed
by visual inspection of the data. 
%The fact that M31 and MW
%GCs occupy the same locus in all panels indicates that the same
%self-enrichment physics operates in the two GC systems.
} 
\label{fig3}
\end{figure}

Resulting [Fe/H] and abundance ratios ([X/Fe]) are plotted as a
function of GC mass in Figure~\ref{fig3}, where GCs are color coded
according to [Fe/H].  Squares (open stars) represent data for M31
(MW) GCs.  The Spearman rank correlation coefficient ($\rho$) and
the probability associated with the null hypothesis (P) for a
correlation between data pairs are shown in the upper left
corner of each panel.  The only quantity correlating
significantly with GC mass is [N/Fe] ($\rho=0.37$, $P=2\times10^{-3}$).
We note that [N/Fe] correlates even more strongly with GC absolute
magnitude, M$_V$, which is a more directly measurable quantity, more
reliable than mass ($\rho=-0.47$, $P=1\times10^{-4}$, not
shown).  The numbers suggest the presence of a weak anti-correlation
between [C/Fe] and mass, although that is not confirmed by visual
inspection of the data.

\begin{figure} 
\includegraphics[scale=0.45]{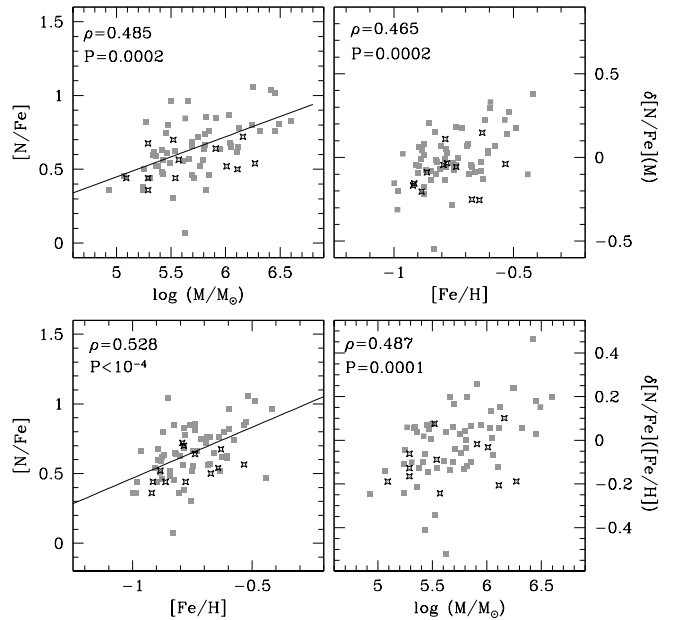}
\caption{The left panels show linear fits to the relations between
[N/Fe] and GC mass (top, repeated from Figure~\ref{fig3}) and [Fe/H]
(bottom).  The right panels plot the residuals from each of these
fits as a function of mass (bottom) and [Fe/H].  Symbols as in
Figure~\ref{fig3}. The residuals from the fit to the [N/Fe]--mass
relation correlate with [Fe/H].  Conversely, the residuals from the
[N/Fe]--[Fe/H] relation correlate with mass.  The correlation between
mass and the residuals from the [N/Fe]--[Fe/H] relation provides
further evidence for the presence of self-enrichment in GCs.  On
the other hand, the correlation between the residuals from the
[N/Fe]--mass relation and [Fe/H] accounts for most of the large
scatter in that relation.  The correlation between [N/Fe] and [Fe/H]
argues for the presence of secondary N-enrichment, which may help
constrain the source of chemical enrichment in GCs.
}
\label{fig4}
\end{figure}

Most of the scatter in the [N/Fe]-mass relation is due to a (stronger)
correlation between [N/Fe] and [Fe/H], which can be perceived in
Figure~\ref{fig3}, where higher [Fe/H] GCs have systematically
higher [N/Fe] at fixed mass.  A similar, although steeper, correlation
was found in early-type galaxies \citep{s07}, which was confirmed
independently by \cite{co13}.  No such correlation can be seen in
solar neighborhood field star data compiled by \cite{s07}, although
this may be due to inhomogeneous abundance data.  Because GCs in the
high [Fe/H] bin span a limited mass range, they only contribute
vertical scatter to the [N/Fe]-mass relation at 5.5$\simless
\log{M/M_\odot} \simless$6.0, leading to a spurious degradation of
the correlation.  Therefore, an accurate assessment of the [N/Fe]-mass
correlation requires restricting the sample to GCs with --1.0 $<$
[Fe/H] $<$ -0.4 (61 GCs), as shown in the upper left panel of
Figure~\ref{fig4}.  The statistical significance of the [N/Fe]-mass 
([N/Fe]-M$_V$) increases in this case, with $\rho$=0.49
($\rho$=--0.51).  Figure~\ref{fig4} also shows linear fits to the
relations of [N/Fe] with mass and [Fe/H] for M31 GCs (left panels).
The residuals from the [N/Fe]--mass relation correlate with
[Fe/H] (upper right), whereas those from the [N/Fe]--[Fe/H] relation
correlate with mass (lower right), providing additional evidence
to the presence of a [N/Fe]-mass relation.

%This can be understood by the following reasoning.  Assuming that
%the slope of the [N/Fe]-mass relation is independent of [Fe/H], if
%our sample contained high (low) GCs with [Fe/H] $>$ --0.4, there
%would be red points in the $\log{M/M_\odot}$ $>$ 6 ($<$ 5.3) and
%[N/Fe] $>$ 1 ($<$ 0.7) region of the mass-[N/Fe] diagram

For a small sample of Galactic GCs, abundance studies of main
sequence stars show a large spread in [N/Fe] and [C/Fe]
\cite[e.g.,][]{br04}.  Because the atmospheres of dwarf stars are
not affected by internal mixing from evolutionary effects, this
spread is usually interpreted as resulting from chemical evolution
within the GCs' stellar populations, with the first generation stars
having an abundance pattern similar to that of field populations
of same [Fe/H], and stars formed subsequently being enhanced in
nitrogen and depleted in carbon.  While observational challenges
prevent abundance measurements in individual M31 GC stars, they are
likely to be present there as well, thus our abundance ratio
measurements should be interpreted as {\it mean} values.  Evidence
from MW GCs suggests that second generation stars represent a
significant, approximately constant, fraction of GC masses today
\citep[e.g.,][]{ca10a,vc11,cs11}.  Therefore mean abundances based
on integrated light spectroscopy provide a reliable indication of
the amount of chemical enrichment undergone by GCs.  Thus, the
correlation between GC mass and mean [N/Fe] is evidence for the
occurence of a feedback-regulated history of star formation and
chemical enrichment in GCs. A somewhat loose correlation between
GC absolute magnitude and the range of [O/Na] in GC stars in a
smaller sample of MW GCs was found before \citep{ca10a}.  However,
the correlation presented in this {\it Letter} is more robust, based
on a larger sample, and for the first time for nitrogen, which is
difficult to measure in large samples of individual stars.

%Scatter in correlations between [N/Fe] and both mass and [Fe/H]
%cannot be accounted by measurement errors alone, given our error
%bars.  Possible sources of intrinsic scatter are moderate amounts
%of mass loss (see discussion below) and varying contributions to
%GC mass by first and second generation stars---which in turn is a
%result of a combination of mass loss and the history of star formation
%in GCs.  

%The fact that both M31
%and MW GCs seem to follow a similar correlation between mean [N/Fe]
%and mass suggests that the chemical enrichment in both GC systems
%is governed by similar physics.

%The latter is further suggested
%by the model prediction in Figure 2a, where [[CHARLIE DESCRIBES
%MODEL BASED ON MW GC CN ABUNDANCES IN 2 LINES]] fits the slope of
%the [N/Fe]-mass relation at constant [Fe/H] very well.

%The strong correlation between [N/Fe] and GC mass in Figure 1
%suggests that M31 GCs have enriched themselves in nitrogen, thus
%providing an important constraint on GC formation models incorporating
%chemical evolution of cluster populations.  

%In common to all recent models of GC formation is the prediction
%that more massive GCs self-enriched more efficiently, because they
%can more easily hold on to stellar ejecta thus enriching the
%intracluster meeting more effectively.  

Average formal uncertainties in abundances (lower right corner of
each panel of Figure~\ref{fig3}) are small, but zero point uncertainties
may be larger.  Systematic effects may affect [N/Fe] values,
especially for [N/Fe] $\gg$ +0.3, where they rely on extrapolations
of the CN$_1$ index sensitivity calculations \citep{s07}.  Further
uncertainties stem from [N/Fe] being derived from the strengths of
CN bands, which are also sensitive to [C/Fe], although the latter
are well constrained by modeling of the C$_2$4668 index.
%However, in Section 1 of the Supplementary Information,
%we discuss model-independent evidence of correlations
%between GC mass and the relative abundances of nitrogen and sodium.
%Moreover, we find a correlation between mass and the NaD
%index, which is sensitive to the abundance of sodium, is further
%evidence to the presence of self-enrichment in the GCs under study.
%For further details, please see Supplementary Information.
%Conversely, other indices such as \Fe, Mg $b$, Ca4227, and C$_2$4668
%present no evidence for a correlation with mass.  (the best case
%being \Fe, for which $\rho$=0.19 and P=0.11).  Since CN$_2$ and
%CN3883 are in different parts of the spectrum and measure different
%transitions of the CN molecule, a correlation between CN concentration
%in the atmospheres of GC stars and GC mass is very likely to be
%present.  Because no similar correlation is found between other
%carbon-sensitive indices (for C$_2$4668 we find $\rho$=0.08 and
%P=0.51), the CN band strength correlation provided further evidence
%to the existence of a correlation between [N/Fe] abundance and GC
%mass.
Most importantly, although model uncertainties may render the
absolute values of [N/Fe] in Figure~\ref{fig3} somewhat uncertain,
the correlation between [N/Fe] and GC mass seems robust.  A substantial
expansion of the data on [C/Fe] and [N/Fe] towards large samples
of individual MW GC stars would help validate the existence
of this trend and uncover its origin.

\begin{figure} 
\includegraphics[scale=0.45]{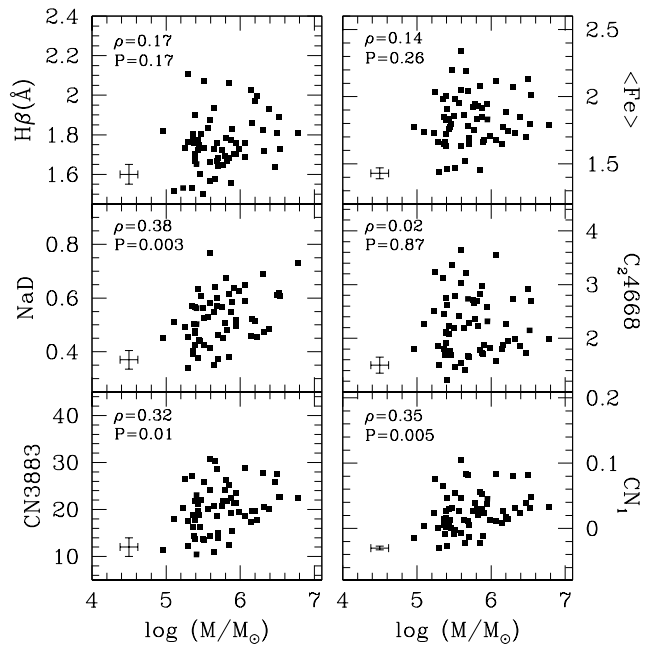}
\caption{Correlations between line indices and GC mass.  The sample
is the same as used in Figures~\ref{fig3} and \ref{fig4}.  Statistically
significant correlations are found for indices sensitive to the abundances
of nitrogen (CN$_1$ and CN3883) and sodium (NaD), whereas indices mainly sensitive
to age ($H\beta$) or the abundances of iron ($<Fe>$) and carbon ($C_24668$)
do not correlate with mass.  Cluster mass correlate more weakly with
line indices than with abundance ratios because the former are sensitive to
other parameters, such as age, metallicity, and horizontal-branch
morphology (especially in the case of CN3883).  These plots
corroborate our conclusion that GCs are self enriched in light
elements, but not in by-produts of explosive nucleosynthesis.
}
\label{fig5}
\end{figure}

Additional, model independent, evidence comes from correlations
between line indices and GC mass.  The correlations are shown in
Figure~\ref{fig5}, where selected line indices are plotted against
GC mass, for the same sample from Figures~\ref{fig3} and \ref{fig4}.
We find statistically significant correlations between GC mass and
CN indices, including Lick CN$_1$ (and CN$_2$, not shown) and CN3883
\citep{dc94}.  The CN bands are sensitive to the abundances of
nitrogen and carbon, but since there is no correlation between
C$_24668$ and GC mass, one concludes that the correlations are
driven by nitrogen.  A statistically significant correlation is
also found between GC mass and the NaD index, which is sensitive
to sodium abundance---the elemental abundance whose variations
constitute most of the evidence for self-enrichment in Galactic GCs
\citep[e.g.,][]{ca10a}.  The absence of correlations between mass
and indices such as $<Fe>$ and Mg $b$ (not shown), which are mostly
sensitive to the abundances of iron and magnesium, respectively,
further corroborates our conclusion that only the abundances of
specific elements, such as nitrogen, and possibly sodium, correlate
with GC mass.

The correlations between GC mass and line indices are unsurprisingly
noisier than that between [N/Fe] and GC mass, since line indices
are not uniquely determined by the abundance of the target elements,
being also affected by other parameters, such as age, metallicity,
HB morphology, and other elemental abundances.  The strength of the
correlation is weakest for CN3883 index, for unclear reasons, which
may be associated with lower S/N blueward of 4000 ${\rm\AA}$ and
possibly the influence of a spread in horizontal branch morphology
at fixed chemical composition and age.  The latter effect should
impact more strongly the violet CN3883 index than its counterpart
at 4170 ${\rm\AA}$.

%Moreover, we find a correlation between mass and the NaD
%index, which is sensitive to the abundance of sodium, is further
%evidence to the presence of self-enrichment in the GCs under study.
%For further details, please see Supplementary Information.
%Conversely, other indices such as \Fe, Mg $b$, Ca4227, and C$_2$4668
%present no evidence for a correlation with mass.  (the best case
%being \Fe, for which $\rho$=0.19 and P=0.11).  Since CN$_2$ and
%CN3883 are in different parts of the spectrum and measure different
%transitions of the CN molecule, a correlation between CN concentration
%in the atmospheres of GC stars and GC mass is very likely to be
%present.  Because no similar correlation is found between other
%carbon-sensitive indices (for C$_2$4668 we find $\rho$=0.08 and
%P=0.51), the CN band strength correlation provided further evidence
%to the existence of a correlation between [N/Fe] abundance and GC
%mass.

%We also note that the NaD index, which is sensitive to Na abundance,
%also correlates with GC mass ($\rho=0.42$ and $P=4\times10^{-4}$)
%and absolute magnitude ($\rho=-0.42$ and $P=3\times10^{-4}$).  

\section{Discussion} \label{summary}

Correlations between cluster mass and mean abundances of N and
(possibly) Na have the following implications for our understanding
of the formation of GCs and the MW halo.  First, they are consistent
with the occurrence of a feedback-regulated history of star formation
and chemical enrichment in M31 GCs.  One would thus expect that the
correlation breaks down at some low mass limit, below which the
gravitational potential would be too shallow to retain chemically
enriched ejecta.  No such effect is detected, which may be due to
the relatively high low-mass limit of our high S/N sample,
M~$>$~10$^{4.8}$M$_\odot$, which is higher than the low mass threshold
for GC self-enrichment in M31, $\sim$10$^{4.5}$M$_\odot$, proposed
by \cite{cs11} on the basis of modeling of pressura balance in
intracluster gas in the early galaxy environment.  Better data are
required for lower mass old GCs, to test that prediction.

Second, many authors \cite[e.g.,][]{re08,vc11,co12} proposed that
MW GCs were as much as 10-100 times more massive in the past, with
most of their stellar masses being lost through two-body relaxation
and tidal stripping during interactions with the Galactic gravitational
potential.  The discovery of CN-strong stars in the MW halo field
led to the suggestion that a substantial fraction of the stellar
halo results from GC dissolution \citep{ma11,ca13}.  However, one
intuitively expects that copious mass loss erase pre-existing
correlations between chemistry and mass, unless, on the one hand,
most mass loss happened {\it before} the bulk of internal chemical
evolution took place in GCs or, on the other, mass loss physics
establishes a smooth relation between primordial and current GC
masses.  On the one hand, numerical simulations by \cite{bm03} show
that most stellar mass loss takes place over timescales of $10^9$~yr,
which implies higher stellar age spreads than allowed by observations
of turnoff stars in both MW and LMC GCs \citep[e.g.,][]{ru13}.  On
the other hand, simulations suggest that mass loss is proportional
to a power of cluster mass \citep{la10}, so that mass loss should
preserve a pre-existing mass-chemical composition relation, provided
it depends more strongly on GC mass than on other variables such
as orbital characteristics and environment.  Yet the latter
dependencies may be important enough to account for the observed
scatter in the [N/Fe]-mass relation.

Third, our results help constrain the nature of stars responsible
for chemical enrichment in GCs.  Four mechanisms have been proposed:
winds from intermediate-mass Asymptotic Giant Branch (AGB) stars
\citep[e.g.,][]{re08,de10}, rapidly rotating massive stars
\citep[e.g.,][]{de07}, mass-transfer binary massive stars
\citep[e.g.,][]{dm09}, and explosive nucleosynthesis \citep[e.g.,][only
for the most massive GCs]{ma09,vc11}.  The correlation between
[N/Fe] and mass, and the absence of correlations for by-products
of explosive nucleosynthesis, such as Mg (and Ca, not shown) suggests
that the enrichment of the intracluster medium was caused by low
energy ejecta.  To first order, the velocity of retained stellar
ejecta should be comparable to typical GC escape velocities, which
range between a few and $\sim$ 100 {\it km s$^{-1}$} \citep{gn02}.
This is in relatively good agreement with typical wind velocities
of AGB stars \citep{du92,lo93} and interacting-binary winds
\citep{dm09}.  Likewise, these escape velocities rule out winds
from massive stars, with typical velocities of order 10$^3$ km
s$^{-1}$ \citep{kp00}, although physical mechanisms such as wind-wind
interactions in binary systems \citep{pl09} and slow winds from
fast-rotating massive stars \citep{de07} were proposed to address
this tension.  While our results for Fe, Mg, and Ca rule out
enrichment by explosive nucleosynthesis it is conceivable that it
plays a role in the most massive GCs in M31, since the theoretical
expectation is for supernovae ejecta to be retained by clusters
with M~$\simgreater~10^7$~M$_\odot$ \citep{ba08}.  Those
clusters are the analogs of massive MW GCs, such as $\omega$ Cen
and M~54, whose stellar populations present a range of [Fe/H]
\citep{pa02,ca10b}.  The existence of a correlation between [N/Fe]
and [Fe/H] is interesting.  \cite{car09} also suggest the existence
of a correlation between the amount of cluster self-enrichment (in
Al) and a combination of [Fe/H] and M$_V$.  There is no evidence
for a correlation between [N/Fe] and [Fe/H] in Galactic field stars,
suggesting that it is an imprint of GC chemical evolution.  This
may be a clue to the source of chemical enrichment in these systems,
possibly indicating the presence of secondary N enrichment.

Finally, from chemical evolution models based on enrichment from
material processed through hot bottom burning in massive AGB stars
\citep[e.g.,][]{de10}, one would expect the [N/Fe]-mass correlation
to be accompanied by a [C/Fe]-mass anti-correlation.  Statistics
suggests the possible presence of a weak anti-correlation
(Figure~\ref{fig3}), but that is not confirmed by visual inspection
of the data.  However, the variation in mean [C/Fe] may be small
enough that it would require more accurate data for its detection.
%---a
%hypothesis that can only be tested through confrontation of more
%accurate independent data with quantitative model predictions.

The existence of a mass-chemical composition relation is a strong
clue to the chemical evolution history of GCs, which can constrain
GC formation theory.  The data discussed here provide a proving
ground to test {\it quantitative} model predictions, including a
chemical evolution prescription based on state-of-the-art yields
and a star formation history grounded on a realistic account of the
physical conditions prevailing in primordial clusters.

\acknowledgments  The authors thank the anonymous referee for a
timely and very helpful review.  RPS thanks Bob Rood ({\it in
memoriam}), Maurizio Salaris, Nate Bastian, and Inger J\o rgensen
for useful discussions.  RPS thanks the support of Gemini Observatory,
where part of this research was conducted.

%, which is operated by
%the Association of Universities for Research in Astronomy, Inc.,
%on behalf of the international Gemini partnership of Argentina,
%Australia, Brazil, Canada, Chile, and the United States of America.
%The hospitality of the Department of Astrophysical Sciences at
%Princeton University, and the Harvard-Smithsonian Center for
%Astrophysics, where this paper was partly conceived, is warmly
%acknowledged.

{}

\end{document}